\documentclass[12pt,preprint]{aastex}
\usepackage{epsf}

\def\eg{{\it e.g.,}}
\def\etal{{\it et al.}}
\def\ie{{\it i.e.,}}

\slugcomment{draft of \today}
\shorttitle{EoS in RHDs}
\shortauthors{Ryu et al.}

\begin{document}

\title{Equation of State in Numerical Relativistic Hydrodynamics}

\author{Dongsu Ryu\altaffilmark{1},
        Indranil Chattopadhyay\altaffilmark{1},
    and Eunwoo Choi\altaffilmark{2}}

\altaffiltext{1}{Department of Astronomy and Space Science, Chungnam National
             University, Daejeon 305-764, Korea:
             ryu@canopus.cnu.ac.kr, indra@canopus.cnu.ac.kr}

\altaffiltext{2}{Department of Physics and Astronomy, Georgia State University,
             P.O. Box 4106, Atlanta, GA 30302-4106, USA:
             echoi@chara.gsu.edu}

\begin{abstract}

Relativistic temperature of gas raises the issue of the equation of
state (EoS) in relativistic hydrodynamics.
We study the EoS for numerical relativistic hydrodynamics, and propose
a new EoS that is simple and yet approximates very closely the EoS of
the single-component perfect gas in relativistic regime.
We also discuss the calculation of primitive variables from
conservative ones for the EoS's considered in the paper, and present
the eigenstructure of relativistic hydrodynamics for a general EoS,
in a way that they can be used to build numerical codes.
Tests with a code based on the Total Variation Diminishing (TVD)
scheme are presented to highlight the differences induced by
different EoS's.

\end{abstract}


\section{Introduction}

Relativistic flows are involved in many high-energy astrophysical
phenomena.
Examples includes relativistic jets from Galactic sources
\citep[see][for reviews]{mir99}, extragalactic jets from AGNs
\citep[see][for reviews]{zen97}, and gamma-ray bursts
\citep[see][for reviews]{mes02}.
In relativistic jets from some Galactic microquasars, intrinsic beam
velocities larger than $0.9c$ are typically required to explain the
observed superluminal motions.
In some powerful extragalactic radio sources, ejections from galactic
nuclei produce true beam velocities of more than $0.98c$.
In the general fireball model of gamma-ray bursts, the internal energy
of gas is converted into the bulk kinetic energy during
expansion and this expansion leads to relativistic outflows with high
bulk Lorentz factors $\ga 100$.
The flow motions in these objects are usually highly nonlinear and
intrinsically complex.
Understanding such relativistic flows is important for correctly
interpreting the observed phenomena, but often studying them is
possible only through numerical simulations.

Numerical codes for special relativistic hydrodynamics (hereafter RHDs)
have been successfully built, based on explicit finite difference upwind
schemes that were originally developed for codes of non-relativistic
hydrodynamics.
These schemes utilize approximate or exact Riemann solvers and the
characteristic decomposition of the hyperbolic system of conservation
equations.
RHD codes based on upwind schemes are able to capture sharp
discontinuities robustly in complex flows, and to describe the
physical solution reliably.
A partial list of such codes includes the followings:
\citet{fk96} based on the van Leer scheme,
\citet{mar96}, \citet{alo99}, and \citet{mpb05} based on the PPM scheme,
\citet{szs01} based on the Godunov scheme, 
\citet{cr05} based on the TVD scheme,
\citet{dol95}, \citet{don98}, \citet{db02}, and \citet{rm05} based
on the ENO scheme,  
and \citet{mb05} based on the HLL scheme. 
Reviews of some numerical approaches and test problems can be found
in \citet{mar03} and \citet{wil03}.

Gas in RHDs is characterized by relativistic fluid speed ($v \sim c$)
and/or relativistic temperature (internal energy much greater than
rest energy), and the latter brings us to the issue of the equation
of state (hereafter EoS) of the gas.
The EoS most commonly used in numerical RHDs, which is designed for
the gas with constant ratio of specific heats, however, is essentially
valid only for the gas of either subrelativistic or ultrarelativistic
temperature.
It is because that is not derived from relativistic kinetic theory.
On the other hand, the EoS of the single-component perfect gas in
relativistic regime can be derived from thermodynamics.
But its form involves the modified Bessel functions \citep[see][]{syn57},
and is too complicated to be implemented in numerical schemes.

In this paper, we study EoS for numerical RHDs.
We first revisit two EoS's previously used in numerical codes,
specifically the one with constant ratio of specific heats, and
the other first used by \citet{math71} and later proposed for 
numerical RHDs by \citet{mpb05}.
We then propose a new EoS which is simple to be implemented in numerical
codes with minimum efforts and minimum computational costs, but at the
same time approximates very closely the EoS of the single-component
perfect gas in relativistic regime.
We also discuss the calculation of primitive variables from conservative
ones for the three EoS's.
Then we present the entire eigenstructure of RHDs for a general EoS,
in a way to be used to build numerical codes.
In order to see the consequence of different EoS's, shock tube tests
performed with a code based on the TVD scheme are presented.
The tests demonstrate the differences in flow structure due to
different EoS's.
Employing a correct EoS should be important to get quantitatively
correct results in problems involving a transition from
non-relativistic temperature to relativistic temperature or vice versa.

This paper is organized as follows.
In sections 2 and 3 we discuss three EoS's and the calculation of primitive
variables from conservative ones for those three.
In sections 4 we present the eigenstructure of RHDs with a general EoS.
In sections 5 and 6 we present a code based on the TVD scheme and shock
tube tests with the code.
Concluding remarks are drawn in section 7.

\section{Relativistic Hydrodynamics}

\subsection{Basic Equations}

The special RHD equations for an ideal fluid can be written in the
laboratory frame of reference as a hyperbolic system of conservation
equations
$$
\frac{\partial D}{\partial t}+\frac{\partial}{\partial x_j}
\left(Dv_j\right) = 0,
\eqno{(1a)}
$$
$$
\frac{\partial M_i}{\partial t}+\frac{\partial}{\partial x_j}
\left(M_iv_j+p\delta_{ij}\right) = 0,
\eqno{(1b)}
$$
$$
\frac{\partial E}{\partial t}+\frac{\partial}{\partial x_j}
\left[\left(E+p\right)v_j\right] = 0,
\eqno{(1c)}
$$
where $D$, $M_i$, and $E$ are the mass density, momentum density, and
total energy density, respectively \citep[see, \eg][]{ll59,wil03}.
The conserved quantities in the laboratory frame are expressed as
$$
D = \Gamma\rho,
\eqno{(2a)}
$$
$$
M_i =\Gamma^2{\rho}hv_i,
\eqno{(2b)}
$$
$$
E = \Gamma^2{\rho}h-p,
\eqno{(2c)}
$$
where $\rho$, $v_i$, $p$, and $h$ are the proper mass density, fluid
three-velocity, isotropic gas pressure and specific enthalpy, respectively,
and the Lorentz factor is given by
$$
\Gamma = \frac{1}{\sqrt{1-v^2}}\qquad{\rm with}\qquad
v^2 = v_x^2+v_y^2+v_z^2.
\eqno{(3)}
$$

In the above, the Latin indices (\eg\ $i$) represents spatial coordinates
and conventional Einstein summation is used.
The speed of light is set to unity ($c\equiv1$) throughout this paper.

\subsection{Equation of State}

The above system of equations is closed with an EoS.
Without loss of generality it is given as
$$
h{\equiv}h(p,\rho).
\eqno{(4)}
$$
Then the general form of polytropic index, $n$, and the general form of
sound speed, $c_s$, respectively can be written as
$$
n = {\rho}\frac{{\partial}h}{{\partial}p}-1, \qquad
c_s^2 = -\frac{{\rho}}{nh}\frac{{\partial}h}{{\partial}{\rho}}.
\eqno{(5)}
$$
In addition we use a variable ${\gamma}_h$ to present the EoS
property conveniently,
$$
{\gamma}_h=\frac{h-1}{\Theta},
\eqno{(6)}
$$
where $\Theta = p/{\rho}$ is a temperature-like variable.

The most commonly used EoS, which is called the ideal EoS (hereafter ID),
is given as
$$
p=(\gamma-1)(e-\rho) \qquad {\rm or} \qquad
h=1+\frac{\gamma\Theta}{\gamma-1}
\eqno{(7)}
$$
with a constant $\gamma$.
Here ${\gamma}=c_p/c_v$ is the ratio of specific heats, and $e$ is
the sum of the internal and rest-mass energy densities in the local frame
and is related to the specific enthalpy as
$$
h=\frac{e+p}{\rho}.
\eqno{(8)}
$$
For ID, ${\gamma}_h=\gamma/(\gamma-1)$ does not depend on $\Theta$.
ID may be correctly applied to the gas of either subrelativistic
temperature with ${\gamma}=5/3$ or ultrarelativistic temperature with
${\gamma}=4/3$.
But ID is rented from non-relativistic thermodynamics, and hence it is
not consistent with relativistic kinetic theory.
For example, we have
$$
n = \frac{1}{\gamma -1}, \qquad
c_s^2 = \frac{\gamma\Theta(\gamma -1)}{\gamma\Theta+\gamma -1}.
\eqno{(9)}
$$
In the high temperature limit, \ie\ ${\Theta}{\rightarrow}\infty$,
and for ${\gamma}>2$, $c_s>1$ \ie\ admits superluminal sound speed.
More importantly, using relativistic kinetic theory \citet{ta48} showed
that the choice of EoS is not arbitrary and has to satisfy the inequality,
$$
(h-\Theta)(h-4\Theta) \geq 1.
\eqno{(10)}
$$
This rules out ID for $\gamma > 4/3$, if applied for $0<\Theta<\infty$.

The correct EoS for the single-component perfect gas in relativistic
regime (hereafter RP) can be derived \citep[see][]{syn57}, and is given as
$$
h=\frac{K_3(1/\Theta)}{K_2(1/\Theta)},
\eqno{(11)}
$$
where $K_2$ and $K_3$ are the modified Bessel functions of the second
kind of order two and three, respectively.
In the non-relativistic temperature limit ($\Theta \rightarrow 0$),
$\gamma_h \rightarrow 5/2$, and in the ultrarelativistic temperature limit
($\Theta \rightarrow \infty$), $\gamma_h \rightarrow 4$.
However, using the above EoS comes with a price of extra computational
costs \citep[][]{fk96}, since the thermodynamics of the fluid is expressed
in terms of the modified Bessel functions.

There have been efforts to find approximate EoS's which are simpler
than RP but more accurate than ID. For example, \citet{szs01} proposed
$$
\Theta=\frac{1}{4}\left(h-\frac{1}{h}\right)
\qquad {\rm or} \qquad
h=2\Theta+\sqrt{4\Theta^2+1}.
\eqno{(12)}
$$
But this EoS does not satisfy either Taub's inequality nor is
consistent with the value of ${\gamma}_h$ in the non-relativistic
temperature limit.

In a recent paper, \citet{mpb05} proposed for numerical RHDs an EoS that
fits RP well.
The EoS, which was first used by \citet{math71}, is given as
$$
\frac{p}{\rho} = \frac{1}{3}\left(\frac{e}{\rho} - \frac{\rho}{e}\right)
\qquad {\rm or} \qquad
h=\frac{5}{2}\Theta+\frac{3}{2}\sqrt{\Theta^2+\frac{4}{9}},
\eqno{(13)}
$$
and is abbreviated as TM following \citet{mpb05}.
With TM the expressions of $n$ and $c_s$ become
$$
n=\frac{3}{2}+\frac{3}{2}\frac{\Theta}{\sqrt{\Theta^2+4/9}},
\qquad
c^2_s=\frac{5\Theta\sqrt{\Theta^2+4/9}+3\Theta^2}
{12\Theta\sqrt{\Theta^2+4/9}+12\Theta^2+2}.
\eqno{(14)}
$$
TM corresponds to the lower bound of Taub's inequality,
\ie\ $(h-\Theta)(h-4\Theta) = 1$.
It produces the right asymptotic values for $\gamma_h$.

In this paper we propose a new EoS, which is a simpler algebraic
function of $\Theta$ and is also a better fit of RP compared to TM.
We abbreviate our proposed EoS as RC and give it by
$$
\frac{p}{e-\rho} = \frac{3p+2\rho}{9p+3\rho}
\qquad {\rm or} \qquad
h=2\frac{6\Theta^2+4\Theta+1}{3\Theta+2}.
\eqno{(15)}
$$
With RC the expressions of $n$ and $c_s$ become
$$
n=3\frac{9\Theta^2+12\Theta+2}{(3\Theta+2)^2},
\qquad
c^2_s=\frac{\Theta(3\Theta+2)(18\Theta^2+24\Theta+5)}
{3(6\Theta^2+4\Theta+1)(9\Theta^2+12\Theta+2)}.
\eqno{(16)}
$$
RC satisfies Taub's inequality, $(h-\Theta)(h-4\Theta) \geq 1$,
for all $\Theta$.
It also produces the right asymptotic values for ${\gamma}_h$.
For both TM and RC, we have correctly $c_s^2 \rightarrow 5\Theta/3$ in
the non-relativistic temperature limit and $c_s^2 \rightarrow 1/3$
in the ultrarelativistic temperature limit, respectively.

In Figure 1, ${\gamma}_h$, $n$, and $c_s$ are plotted as a function of
$\Theta$ to compare TM and RC to RP as well as ID.
One can see that RC is a better fit of RP than TM with
$$
\frac{\left|h_{\rm TM}-h_{\rm RP}\right|}{h_{\rm RP}} \la 2\%,
\qquad
\frac{\left|h_{\rm RC}-h_{\rm RP}\right|}{h_{\rm RP}} \la 0.8\%.
\eqno{(17)}
$$
It is to be remembered that both ${\gamma}_h$ and $n$ are independent
of $\Theta$, if ID is used.

\section{Calculation of Primitive Variables}

The RHD equations evolve the conserved quantities, $D$, $M_i$ and $E$,
but we need to know the values of the primitive variables, $\rho$,
$v_i$, $p$, to solve the equations numerically.
The primitive variables can be calculated by inverting the
equations (2a--2c).
The equations (2a--2c) explicitly include $h$, and here we discuss the
inversion for the EoS's discussed in section 2.2, that is, ID, TM, and RC.

\subsection{ID}

\citet{sch93} showed that the equations (2a--2c) with the EoS in (7)
reduce to a single quartic equation for $v$
$$
v^4+b_1v^3+b_2v^2+b_3v+b_4=0,
\eqno{(18)}
$$
where
$$
b_1=-\frac{2{\gamma}({\gamma}-1)ME}{({\gamma}-1)^2(M^2+D^2)},
\qquad
b_2=\frac{{\gamma}^2E^2+2({\gamma}-1)M^2-({\gamma}-1)^2D^2}
{({\gamma}-1)^2(M^2+D^2)},
\eqno{(19a)}
$$
$$
b_3=-\frac{2{\gamma}ME}{({\gamma}-1)^2(M^2+D^2)},
\qquad
b_4=\frac{M^2}{({\gamma}-1)^2(M^2+D^2)},
\eqno{(19b)}
$$
and $M=\sqrt{M^2_x+M^2_y+M^2_z}$.
The quartic equation (18) can be solved numerically or analytically.
In \citet{cr05} the analytical solution was used for the very first
time, though the exact nature of the solution was not presented. 

The general form of analytical roots for quartic equations can be
found in \citet{abr72} or on webs such as
``http://mathworld.wolfram.com/QuarticEquation.html''.
One may even use softwares such as Mathematica or Maxima to find
the roots.
We found that out of the four roots of the quartic equation (18),
two are complex and two are real.
The two real roots are
$$
z_1=\frac{-B+\sqrt{B^2-4C}}{2},
\qquad
z_2=\frac{-B-\sqrt{B^2-4C}}{2},
\eqno{(20)}
$$
where
$$
B=\frac{1}{2}(b_1+\sqrt{b^2_1-4b_2+4x_1}),
\qquad
C=\frac{1}{2}(x_1-\sqrt{x^2_1-4b_4}),
\eqno{(21a)}
$$
$$
x_1=(R+T^{\frac{1}{2}})^{\frac{1}{3}}
+(R-T^{\frac{1}{2}})^{\frac{1}{3}}-\frac{a_1}{3},
\eqno{(21b)}
$$
$$
R=\frac{9a_1a_2-27a_3-2a^3_1}{54},
\qquad
S=\frac{3a_2-a^2_1}{9},
\qquad
T=R^2+S^3,
\eqno{(21c)}
$$
$$a_1=-b_2,
\qquad
a_2=b_1b_3-4b_4,
\qquad
a_3=4b_2b_4-b^2_3-b^2_1b_4.
\eqno{(21d)}
$$
Among the two real roots, the first one is the solution that satisfies
the upper and lower limits imposed by \citet{sch93}, thus $v=z_1$.
Once $v$ is found, the quantities $\rho$, $v_i$, $p$, are calculated by
$$
\rho=\frac{D}{\Gamma},
\eqno{(22a)}
$$
$$
v_x=\frac{M_x}{M}v,
\qquad
v_y = \frac{M_y}{M}v,
\qquad
v_z = \frac{M_z}{M}v,
\eqno{(22b)}
$$
$$
p=(\gamma-1)[(E-M_xv_x-M_yv_y-M_zv_z)-{\rho}].
\eqno{(22c)}
$$

\subsection{TM}

Combining the equations (2a--2c) with the EoS in (13), we get a cubic
equation for $W=\Gamma^2-1$
$$
W^3+c_1W^2+c_2W+c_3=0,
\eqno{(23)}
$$
where
$$
c_1=\frac{(E^2+M^2)[4(E^2+M^2)-(M^2+D^2)]-14M^2E^2}{2(E^2-M^2)^2},
\eqno{(24a)}
$$
$$ 
c_2=\frac{[4(E^2+M^2)-(M^2+D^2)]^2-57M^2E^2}{16(E^2-M^2)^2},
\eqno{(24b)}
$$
$$ 
c_3=-\frac{9M^2E^2}{16(E^2-M^2)^2}.
\eqno{(24c)}
$$

Cubic equations admit analytical solutions simpler than quartic
equations \citep[see also][]{abr72}.
We found that out of the three roots of the cubic equation (23),
two are unphysical giving $\Gamma<1$, and only one gives the physical
solution, which is
$$
W=2{\sqrt{-J}} \cos\left(\frac{\iota}{3}\right)-\frac{c_1}{3},
\eqno{(25)}
$$
where
$$J=\frac{3c_2-c^2_1}{9},
\qquad
\cos{\iota}=\frac{H}{\sqrt{-J^3}},
\qquad
H=\frac{9c_1c_2-27c_3-2c^3_1}{54}.
\eqno{(25)}
$$
Then the fluid speed is calculated by
$$
v=\frac{W}{\sqrt{W^2+1}},
\eqno{(27)}
$$
and the quantities $\rho$, $v_i$, $p$, are calculated by
$$
\rho=\frac{D}{\Gamma}.
\eqno{(28a)}
$$
$$
v_x=\frac{M_x}{M}v,
\qquad
v_y = \frac{M_y}{M}v,
\qquad
v_z = \frac{M_z}{M}v,
\eqno{(28b)}
$$
$$
p=\frac{(E-M_xv_x-M_yv_y-M_zv_z)^2-{\rho}^2}{3(E-M_xv_x-M_yv_y-M_zv_z)}.
\eqno{(28c)}
$$

\subsection{RC}

Combining the equations (2a--2c) with the EoS in (15), we get
$$
M\sqrt{\Gamma^2-1}\left[3E\Gamma(8\Gamma^2-1)+2D(1-4\Gamma^2)\right]
$$
$$
= 3\Gamma^2\left[4(M^2+E^2)\Gamma^2-(M^2+4E^2)\right]
-2D(4E\Gamma-D)(\Gamma^2-1).
\eqno{(29)}
$$
Further simplification reduces it into an equation of $8^{\rm th}$
power in $\Gamma$.

Although the equation (29) has to be solved numerically, it behaves
very well.
We first analyzed the nature of the roots with a root-finding routine
in the IMSL library.
As noted by \citet{sch93}, the physically meaningful solution should be
between the upper limit,
$\Gamma_u$,
$$
\Gamma_u=\frac{1}{\sqrt{1-v^2_u}}
\qquad {\rm with} \qquad
v_u=\frac{M}{E},
\eqno{(30)}
$$
and the lower limit, $\Gamma_l$, that is derived inserting $D=0$ into
equation (29):
$$
16(M^2-E^2)^2\Gamma_l^6-8(M^2-E^2)(M^2-4E^2)\Gamma_l^4
+(M^4-9M^2E^2+16E^4)\Gamma_l^2+M^2E^2=0
\eqno{(31)}
$$
(a cubic equation of $\Gamma_l^2$).
Out of the eight roots of the equation (29), four are complex and
four are real.
Out of the four real roots, two are negative and two are positive.
And out of the two real and positive roots, one is always larger
than $\Gamma_u$, and the other is between $\Gamma_l$ and
$\Gamma_u$ and so is the physical solution.

Inside RHD codes the physical solution of equation (29) can be easily
calculated by the Newton-Raphson method.
With an initial guess $\Gamma=\Gamma_l$ or any value smaller than it
including 1, iteration can be proceeded upwards.
Since the equation is extremely well-behaved, the iteration converges
within a few steps.
Once $\Gamma$ is known, the fluid speed is calculated by
$$
v=\frac{\sqrt{\Gamma^2-1}}{\Gamma},
\eqno{(32)}
$$
and the quantities $\rho$, $v_i$, $p$, are calculated by
$$
\rho=\frac{D}{\Gamma}.
\eqno{(33a)}
$$
$$
v_x=\frac{M_x}{M}v,
\qquad
v_y = \frac{M_y}{M}v,\qquad
v_z = \frac{M_z}{M}v
\eqno{(33b)}
$$
$$
p=\frac{(E-M_iv_i)-2\rho
+\left[(E-M_iv_i)^2+4\rho(E-M_iv_i)-4\rho^2\right]^\frac{1}{2}}{6},
\eqno{(33c)}
$$
where
$$
M_iv_i=M_xv_x+M_yv_y+M_zv_z.
\eqno{(34)}
$$

\section{Eigenvalues and Eigenvectors}

In building a code based on the Roe-type schemes such as the TVD and
ENO schemes that solves a hyperbolic system of conservation equations,
the eigenstructure (eigenvalues and eigenvectors of the Jacobian matrix)
is required.
The Eigenstructure for RHDs was previously described, for instance,
in \citet{don98}.
However, with the parameter vector different from that of \citet{don98},
the eigenvectors become different.
Here we present our complete set of eigenvalues and eigenvectors
without assuming any particular form of EoS.

Equations (1a)--(1c) can be written as
$$
\frac{\partial \vec{q}}{\partial t}
+\frac{\partial \vec{F}_j}{\partial x_j} = 0
\eqno{(35)}
$$
with the state and flux vectors
$$
\vec{q} = \left[\matrix{D\cr M_i\cr E}\right], \qquad
\vec{F}_j = \left[\matrix{Dv_j\cr M_iv_j+p\delta_{ij}\cr
\left(E+p\right)v_j}\right],
\eqno{(36)}
$$
or as
$$
\frac{\partial\vec{q}}{\partial t}
+A_j\frac{\partial\vec{q}}{\partial x_j} = 0, \qquad
A_j = \frac{\partial\vec{F}_j}{\partial\vec{q}}.
\eqno{(37)}
$$
Here $A_j$ is the $5\times5$ Jacobian matrix composed with the
state and flux vectors.
The construction of the matrix $A_j$ can be simplified by introducing
a parameter vector, $\vec{u}$, as
$$
A_j = \frac{\partial\vec{F}_j}{\partial\vec{u}}
\frac{\partial\vec{u}}{\partial\vec{q}}.
\eqno{(38)}
$$
We choose the vector made of primitive variables as the parameter
vector
$$
\vec{u} = \left[\matrix{\rho\cr v_i\cr p}\right].
\eqno{(39)}
$$

\subsection{One Velocity Component}

The eigenstructure is simplified if only a single component of velocity
is chosen, \ie\ $v=v_x$.
In principle it can be reduced from that with three components of
velocity in the next subsection.
Nevertheless we present it, for the case that the simpler eigenstructure
with one velocity component can be used.

The explicit form of the Jacobian matrix, $A$, is presented in Appendix A.
The eigenvalues of $A$ are,
$$
a_-=\frac{v-c_s}{1-c_sv},
\qquad
a_o=v,
\qquad
a_+=\frac{v+c_s}{1+c_sv}.
\eqno{(40)}
$$ 
The right eigenvectors are
$$
\vec{R}_-=\left[\matrix{1\cr \Gamma h(v-c_s)\cr \Gamma h(1-c_sv)}\right],
\qquad
\vec{R}_0=\left[\matrix{1\cr \Gamma hv(1-nc^2_s)\cr \Gamma h(1-nc^2_s)}\right],
\qquad
\vec{R}_+=\left[\matrix{1\cr \Gamma h(v+c_s)\cr \Gamma h(1+c_sv)}\right].
\eqno{(41)}
$$
and the left eigenvectors are
$$
\vec{L}_-=-\frac{1}{2hnc^2_s}\left[h(1-nc^2_s),~~ \Gamma (v+nc_s),~~
-\Gamma (1+nc_sv) \right],
\eqno{(42a)}
$$
$$
\vec{L}_0=\frac{1}{hnc^2_s}\left[h,~~ \Gamma v,~~ -\Gamma \right],
\eqno{(42b)}
$$
$$
\vec{L}_+=-\frac{1}{2hnc^2_s}\left[h(1-nc^2_s),~~ \Gamma (v-nc_s),~~
-\Gamma (1-nc_sv) \right].
\eqno{(42c)}
$$
Here $n$ and $c_s$ are given in equation (5).

\subsection{Three Velocity Components}

The $x$-component of the Jacobian matrix, $A_x$, when all the three
components of velocity are considered, is presented in Appendix B.
The eigenvalues of $A_x$ are
$$
a_1 = \frac{\left(1-c_s^2\right)v_x-c_s/\Gamma\cdot\sqrt{Q}}{1-c_s^2v^2},
\eqno{(43a)}
$$
$$
a_2 = v_x,
\eqno{(43b)}
$$
$$
a_3 = v_x,
\eqno{(43c)}
$$
$$
a_4 = v_x,
\eqno{(43d)}
$$
$$
a_5 = \frac{\left(1-c_s^2\right)v_x+c_s/\Gamma\cdot\sqrt{Q}}{1-c_s^2v^2},
\eqno{(43e)}
$$
where $Q=1-v^2_x-c^2_s(v^2_y+v^2_z)$.
The eigenvalues represent the five characteristic speeds associated with
two sound wave modes ($a_1$ and $a_5$) and three entropy modes ($a_2$,
$a_3$, and $a_4$).
A remarkable feature is that the eigenvalues do not explicitly depend on
$h$ and $n$, but only on $v_i$ and $c_s$.
Hence the eigenvalues are the same regardless of the choice of EoS once
the sound speed is defined properly.

The corresponding right eigenvectors ($A_x \vec{R} = a \vec{R}$), however,
depends explicitly on $h$ and $n$, and the complete set is given by
$$
\vec{R}_1 = \left[\frac{1-a_1v_x}{\Gamma},~~a_1h(1-v^2_x),~~
h(1-a_1v_x)v_y,~~h(1-a_1v_x)v_z,~~h(1-v^2_x)\right]^\mathrm{T},
\eqno{(44a)}
$$
$$
\vec{R}_2 = {\tilde X}\left[X_1,~~X_2,~~X_3,~~X_4,~~X_5\right]^\mathrm{T},
\eqno{(44b)}
$$
$$
\vec{R}_3 = \frac{1}{1-v^2_x}\left[\frac{v_y}{\Gamma h},~~2v_xv_y,~~
1-v^2_x+v^2_y,~~v_yv_z,~~2v_y\right]^\mathrm{T},
\eqno{(44c)}
$$
$$
\vec{R}_4 = \frac{1}{1-v^2_x}\left[\frac{v_z}{\Gamma h},~~2v_xv_z,~~
v_yv_z,~~1-v^2_x+v^2_z,~~2v_z\right]^\mathrm{T},
\eqno{(44d)}
$$
$$
\vec{R}_5 = \left[\frac{1-a_5v_x}{\Gamma},~~a_5h(1-v^2_x),~~
h(1-a_5v_x)v_y,~~h(1-a_5v_x)v_z,~~h(1-v^2_x)\right]^\mathrm{T},
\eqno{(44e)}
$$
where
$$
X_1=\frac{nc^2_s(v^2_y+v^2_z)+(1-v^2_x)}{\Gamma h},
\eqno{(45a)}
$$
$$
X_2=\left[2nc^2_s(v^2_y+v^2_z)+(1-nc^2_s)(1-v^2_x)\right]v_x,
\eqno{(45b)}
$$
$$
X_3=\left[nc^2_s(v^2_y+v^2_z)+(1-v^2_x)\right]v_y,
\eqno{(45c)}
$$
$$
X_4=\left[nc^2_s(v^2_y+v^2_z)+(1-v^2_x)\right]v_z,
\eqno{(45d)}
$$
$$
X_5=2nc^2_s(v^2_y+v^2_z)+(1-nc^2_s)(1-v^2_x).
\eqno{(45e)}
$$
$$
{\tilde X}=\frac{{\Gamma}^2}{nc^2_s(1-v^2_x)},
\eqno{(45f)}
$$

The complete set of the left eigenvectors ($\vec{L} A_x = a \vec{L}$),
which are orthonormal to the right eigenvectors, is
$$
\vec{L}_1 = \frac{1}{\tilde Y}_1
\left[Y_{11},~~Y_{12},~~Y_{13},~~Y_{13},~~Y_{15}\right],
\eqno{(46a)}
$$
$$
\vec{L}_2 = \left[\frac{h}{\Gamma},~~v_x,~~v_y,~~v_z,~~-1\right],
\eqno{(46b)}
$$
$$
\vec{L}_3 = \left[-{\Gamma}hv_y,~~0,~~1,~~0,~~0\right],
\eqno{(46c)}
$$
$$
\vec{L}_4 = \left[-{\Gamma}hv_z,~~0,~~0,~~1,~~0\right],
\eqno{(46d)}
$$
$$
\vec{L}_5 = \frac{1}{\tilde Y}_5
\left[Y_{51},~~Y_{52},~~Y_{53},~~Y_{53},~~Y_{55}\right],
\eqno{(46e)}
$$
where
$$
Y_{i1}=-\frac{h}{\Gamma}(1-a_iv_x)(1-nc^2_s),
\eqno{(47a)}
$$
$$
Y_{i2}=na_i(1-c^2_sv^2)+a_i(1+nc^2_s)v^2_x-(1+n)v_x,
\eqno{(47b)}
$$
$$
Y_{i3}=-(1+nc^2_s)(1-a_iv_x)v_y,
\eqno{(47c)}
$$
$$
Y_{i4}=-(1+nc^2_s)(1-a_iv_x)v_z,
\eqno{(47d)}
$$
$$
Y_{i5}=(1+nc^2_sv^2)+(1-c^2_s)nv^2_x-a_i(1+n)v_x,
\eqno{(47e)}
$$
$$
{\tilde Y}_i=hn\left[(a_i-v_x)^2Q+\frac{c^2_s}{\Gamma ^2}\right],
\eqno{(47f)}
$$
and index $i=1,\ 5$.

We note that with three degenerate modes that have same eigenvalues,
$a_2 = a_3 = a_4$, we have a freedom to write down the right and left
eigenvectors in a variety of different forms.
We chose to present the ones that produce the best results with
the TVD code described next.

\section{One-Dimensional Functioning Code}

To be used for demonstration of the differences in flow structure due to
different EoS's, a one-dimensional functioning code based on the Total
Variation Diminishing (TVD) scheme was built.
The code utilizes the eigenvalues and eigenvectors given in the previous
section, and can employ arbitrary EoS's including those in section 2.2.

\subsection{The TVD Scheme}

The TVD scheme, originally developed by \citet{har83}, is an Eulerian,
finite-difference scheme with second-order accuracy in space and time.
The second-order accuracy in time is achieved by modifying numerical
flux using the quantities in five grid cells
\citep[see below and][for details]{har83}.
The scheme  is basically identical to that previously used in
\citet{ryu93} and \citet{cr05}.
But for completeness, the procedure is concisely shown here.

The state vector $\vec{q}_i^n$ at the cell center $i$ at the time step
$n$ is updated by calculating the modified flux vector
$\bar{\vec{f}}_{x,i\pm1/2}$ along the $x$-direction at the cell interface
$i\pm1/2$ as follows:
$$
L_x\vec{q}_i^n = \vec{q}_i^n-\frac{\Delta t^n}{\Delta x}
\left(\bar{\vec{f}}_{x,i+1/2}-\bar{\vec{f}}_{x,i-1/2}\right),
\eqno{(48)}
$$
$$
\bar{\vec{f}}_{x,i+1/2} = \frac{1}{2}\left[\vec{F}_x(\vec{q}_i^n)
+\vec{F}_x(\vec{q}_{i+1}^n)\right]-\frac{\Delta x}{2\Delta t^n}
\sum_{k=1}^5\beta_{k,i+1/2}\vec{R}_{k,i+1/2}^n,
\eqno{(49)}
$$
$$
\beta_{k,i+1/2} = Q_k(\frac{\Delta t^n}{\Delta x}a_{k,i+1/2}^n
+\gamma_{k,i+1/2})\alpha_{k,i+1/2}-\left(g_{k,i}+g_{k,i+1}\right),
\eqno{(50)}
$$
$$
\gamma_{k,i+1/2} = \left\{
\begin{array}{lcl}
\left(g_{k,i+1}-g_{k,i}\right)/\alpha_{k,i+1/2} & \mathrm{for} &
\alpha_{k,i+1/2}\neq0, \\
0 & \mathrm{for} & \alpha_{k,i+1/2}=0,
\end{array}
\right.
\eqno{(51)}
$$
$$
g_{k,i} = \mathrm{sign}(\tilde{g}_{k,i+1/2})\mathrm{max}
\{0,\mathrm{min}[|\tilde{g}_{k,i+1/2}|,
\mathrm{sign}(\tilde{g}_{k,i+1/2})\tilde{g}_{k,i-1/2}]\},
\eqno{(52a)}
$$
$$
g_{k,i} = \mathrm{sign}(\tilde{g}_{k,i+1/2})\mathrm{max}
\{0,\mathrm{min}[\frac{1}{2}(|\tilde{g}_{k,i+1/2}|
+\mathrm{sign}(\tilde{g}_{k,i+1/2})\tilde{g}_{k,i-1/2}),
$$
$$
2|\tilde{g}_{k,i+1/2}|,2\mathrm{sign}(\tilde{g}_{k,i+1/2})
\tilde{g}_{k,i-1/2}]\},
\eqno{(52b)}
$$
$$
g_{k,i} = \mathrm{sign}(\tilde{g}_{k,i+1/2})\mathrm{max}
\{0,\mathrm{min}[|\tilde{g}_{k,i+1/2}|,
2\mathrm{sign}(\tilde{g}_{k,i+1/2})\tilde{g}_{k,i-1/2}],
$$
$$
\mathrm{min}[2|\tilde{g}_{k,i+1/2}|,
\mathrm{sign}(\tilde{g}_{k,i+1/2})\tilde{g}_{k,i-1/2}]\},
\eqno{(52c)}
$$
$$
\tilde{g}_{k,i+1/2} = \frac{1}{2}\left[Q_k(\frac{\Delta t^n}{\Delta x}
a_{k,i+1/2}^n)-\left(\frac{\Delta t^n}{\Delta x}a_{k,i+1/2}^n\right)^2
\right]\alpha_{k,i+1/2},
\eqno{(53)}
$$
$$
\alpha_{k,i+1/2} = \vec{L}_{k,i+1/2}^n\cdot\left(\vec{q}_{i+1}^n
-\vec{q}_i^n\right),
\eqno{(54)}
$$
$$
Q_k(x) = \left\{
\begin{array}{lcl}
x^2/4\varepsilon_k+\varepsilon_k & \mathrm{for} & |x|<2\varepsilon_k, \\
|x| & \mathrm{for} & |x|\geq2\varepsilon_k.
\end{array}
\right.
\eqno{(55)}
$$
Here, $k=1$ to 5 stand for the five characteristic modes.
The internal parameters $\varepsilon_k$'s implicitly control numerical
viscosity, and are defined for $0 \leq \varepsilon_k < 0.5$.
The flux limiters in equations (52a)--(52c) are the min-mod, monotonized
central difference, and superbee limiters, respectively, a partial list
of the limiters that are consistent with the TVD scheme, and one of them
has to be employed.

\subsection{Quantities at Cell Interfaces}

To calculate the fluxes we need to define the local quantities
at the cell interfaces, $i+1/2$.
The TVD scheme originally used the Roe's linearizion technique
\citep{roe81} for it.
Although it is possible to implement this linearizion technique in
the relativistic domain in a computationally feasible way
\citep[see][]{eul95}, there is unlikely to be a significant advantage
from the computational point of view.
Instead, we simply use the algebraic averages of quantities at two
adjacent cell centers to define the fluid three-velocity and specific
enthalpy at the cell interfaces:
$$
v_{x,i+1/2} = \frac{v_{x,i}+v_{x,i+1}}{2},
\qquad
v_{y,i+1/2} = \frac{v_{y,i}+v_{y,i+1}}{2},
\qquad
v_{z,i+1/2} = \frac{v_{z,i}+v_{z,i+1}}{2},
\eqno{(56)}
$$
$$
h_{i+1/2} = \frac{h_i+h_{i+1}}{2}.
\eqno{(57)}
$$

Defining $n$ and $c_s$ for the calculation of eigenvalues and eigenvectors
at the cell interfaces depends on EoS.
For ID, $n$ is constant and
$$
c_{s,i+1/2} = \left(\frac{h_{i+1/2}-1}{nh_{i+1/2}}\right)^{1/2}.
\eqno{(58)}
$$
For TM, we first compute from equation (13)
$$
{\Theta}_{i+1/2}=\frac{5h_{i+1/2}-\sqrt{9h^2_{i+1/2}+16}}{8},
\eqno{(59)}
$$
then define $n_{i+1/2}$ and $c_{s,i+1/2}$ according to equation (14).
For RC, we first compute from equation (15)
$$
{\Theta}_{i+1/2}=\frac{3h_{i+1/2}-8+\sqrt{9h_{i+1/2}^2+48h_{i+1/2}-32}}
{24}
\eqno{(60)}
$$
then define $n_{i+1/2}$ and $c_{s,i+1/2}$ according to equation (16).

\section{Numerical Tests}

The differences induced by different EoS's are illustrated through
a series of shock tube tests performed with the code described in
the previous section.
We use the tests used in previous works \citep[\eg][]{mar03,mpb05},
instead of inventing our own.
Two sets are considered, one being purely one-dimensional with only
the velocity component parallel to structure propagation, and the
other with transverse velocity component.

For the first set with parallel velocity component only, two tests
are presented:\\
P1: $\rho_L=10$, $\rho_R=1$, $p_L=13.3$, $p_R=10^{-6}$,  and
$v_{p,L} = v_{p,R} = 0$ initially, and $t_{\rm end} = 0.45$,\\
P2: $\rho_L = \rho_R=1$, $p_L=10^3$, $p_R=10^{-2}$, and
$v_{p,L} = v_{p,R} = 0$ initially, and $t_{\rm end} = 0.4$.\\
The box covers the region of $0 \le x \le 1$.
Here the subscripts $L$ and $R$ denote the quantities in the left
and right states of the initial discontinuity at $x=0.5$, and
$t_{\rm end}$ is the time when the solutions are presented.
These two tests have been extensively used for tests of RHD codes
with the ID EoS \citep[see][]{mar03}, and the analytic solutions
were described in \citet{mar94}.

Figures 2 and 3 show the numerical solutions with RC and TM, and the
analytic solutions with ID and $\gamma = 5/3$ and $4/3$.
The numerical solutions with RC and TM were obtained using the version
of the TVD code having one velocity component (see section 4.1), and
the analytic solutions with ID comes from the routine described
in \citet{mar94}.
The numerical solutions with ID are almost indistinguishable from the
analytic solutions, once they are calculated.

The ID solutions with $\gamma = 4/3$ and $5/3$ show noticeable
differences.
The density shell between the contact discontinuity (hereafter CD)
and the shock becomes thinner and taller with smaller $\gamma$,
because the post shock pressure is lower and so is the shock
propagation speed.
The rarefaction wave is less elongated with $\gamma = 4/3$, because
the sound speed is lower.
Those solutions with ID are also clearly different from the solutions
obtained with RC and TM.
The ID solution with $\gamma = 4/3$ better approximates the solutions
with RC and TM in the left region of the CD, because the flow has
relativistic temperature of $\Theta \ga 1$ there.
The difference is, however, obvious in the shell between the CD and
the shock, because $\Theta \sim 1$ there.
On the other hand, the solutions obtained with RC and TM look very
much alike.
It reflects the similarity in the distributions of specific enthalpy
in equations (13) and (15). 
Yet there is a noticeable difference, especially in the shell between
the CD and the shock, and the difference in density reaches up to
$\sim 5\%$.

For the second set with transverse velocity component, four tests,
where different transverse velocities were added to the test P2, are
presented:\\
T1: initially $v_{t,R}=0.99$ to the right state, $t_{\rm end} = 0.45$,\\
T2: initially $v_{t,L}=0.9$ to the left state, $t_{\rm end} = 0.55$,\\
T3: initially $v_{t,L}=v_{t,R}=0.99$ to the left and right states,
$t_{\rm end} = 0.18$,\\
T4: initially $v_{t,L}=0.9$ and $v_{t,R}=0.99$ to the left and right states,
$t_{\rm end} = 0.75$.\\
The notations are the same ones used in P1 and P2.
These are subsets of the tests originally suggested by \citet{pmm00}
with the ID EoS and later used by \citet{mpb05}.

Figures 4, 5, 6 and 7 show the numerical solutions with RC and TM
and the analytic solutions with ID and $\gamma = 5/3$ and $4/3$.
The numerical solutions with RC and TM were obtained using the version
of the TVD code having three velocity components (see section 4.2), and
the analytic solutions with ID comes from the routine described
in \citet{pmm00}.

Again the ID solutions with $\gamma = 4/3$ and $5/3$ show noticeable
differences.
Especially with transverse velocity initially on the left side of the
initial discontinuity (Figure 5, 6 and 7), the parallel velocity reaches
lower values, while the transverse velocity achieves higher values, with
higher $\gamma = 5/3$ in the region to the left of the CD.
As a result, the density shell between the CD and the shock has
propagated less.
As in the P tests, the solutions with ID are clearly different from the
solutions obtained with RC and TM, most noticeably in the shell between
the CD and the shock.
The solutions with RC and TM look very much alike with differences in the
density in the shell between the CD and the shock of about $\sim 5\%$.

We note that this paper is intended to focus on the EoS in numerical
RHDs, not intended to present the performance of the code.
Hence, one-dimensional tests of high resolution (with $2^{16}$ grid
cells for the P tests and $2^{17}$ grid cells the T tests) are presented
to manifest the difference induced by different EoS's.
The performance of the code such as capturing of shocks and CDs will be
discussed elsewhere.

\section{Summary and Discussion}

The conservation equations for both Newtonian hydrodynamics and RHDs
are strictly hyperbolic, rendering the apt use of upwind schemes for
numerical codes.
The actual implementation to RHDs is, however, complicated, partly
due to EoS.
In this paper we study three EoS's for numerical RHDs, two being
previously used and the other being newly proposed.
The new EoS is simple and yet approximates the enthalpy of
single-component perfect gas in relativistic regime with accuracy
better than $0.8\%$.
Then we discuss the calculation of primitive variables from
conservative ones for the EoS's considered.
We also present the eigenvalues and eigenvectors of RHDs for a general
EoS, in a way that they are ready to be used to build numerical codes
based on the Roe-type schemes such as the TVD and ENO schemes.
Finally we present numerical tests to show the differences in flow
structure due to different EoS's

The most commonly used, ideal EoS, can be used for the gas of
entirely non-relativistic temperature ($\Theta \ll 1$) with
$\gamma=5/3$ or for the gas of entirely ultrarelativistic temperature
($\Theta \gg 1$) with $\gamma=4/3$.
However, if the transition from non-relativistic to relativistic or
vice versa with $\Theta\sim 0.1 - 1$ is involved, the ideal EoS produces
incorrect results and its use should be avoided.
The EoS proposed by \citet{mpb05}, TM, produces reasonably correct
results with error of a few percent at most.
The most preferable advantage of using TM is that the calculation of
primitive variables admits analytic solutions, thereby making
its implementation easy.
The newly suggested EoS, RC, which approximates the EoS of the
relativistic perfect gas, RP, most accurately, produces
thermodynamically the most accurate results.
At the same time it is simple enough to be implemented to numerical
codes with minimum efforts and minimum computational costs.
With RC the primitive variables should be calculated numerically by
an iteration method such as the Newton-Raphson method.
However, the equation for the calculation of primitive variables
behaves extremely well, so the iteration converges in a few step
without any trouble.

In Galactic and extragalactic jets and gamma-ray bursts, as the flows
travel relativistic fluid speeds ($v \sim 1$ but $\Theta \ll 1$),
they would hit the surrounding media.
Then shocks are produced and the gas can be heated up to $\Theta \ga 1$.
These kind of transitions, continuous or discontinuous, between
relativistic bulk speeds and relativistic temperatures are intrinsic
in astrophysical relativistic flows, and so a correct EoS is required
to simulate the flows correctly.
The correctness as well as the simplicity make RC suitable for
astrophysical applications like these.

\acknowledgments{
The work of DR and IC was supported by the KOSEF grant
R01-2004-000-10005-0.
The work of EC was supported by RPE funds to PEGA at GSU.}

\appendix

\section{Jacobian Matrix with One Velocity Component}

$$
A=\frac{1}{N}\left(\begin{array}{ccc}
A_{11}&A_{12}&A_{13} \\
A_{21}&A_{22}&A_{23} \\
0     &A_{32}&0      \\
\end{array} \right)
\eqno{(A.1)}
$$
$$
A_{11}=v^2hn(1-c^2_s)+\frac{vh}{\Gamma^2}
$$
$$
A_{12}=- \frac{1}{\Gamma^3}+\frac{1+n}{\Gamma}
$$
$$
A_{13}=-\frac{v(1+n)}{\Gamma}
$$
$$
A_{21}=-\frac{h^2}{\Gamma^3}(1-nc^2_s)
$$
$$
A_{22}=-\frac{vh}{\Gamma^2}(1-nc^2_s)+2vhn(1-c^2_s)
$$
$$
A_{23}=-v^2hn(1-c^2_s)+\frac{h}{\Gamma^2}
$$
$$
A_{32}=hn(1-c^2_sv^2)
$$
$$
N=hn(1-c^2_sv^2)
$$

\section{Jacobian Matrix with Three Velocity Components}

$$
A_x=\frac{1}{N}\left(\begin{array}{ccccc}
A_{11}&A_{12}&A_{13}&A_{14}&A_{15} \\
A_{21}&A_{22}&A_{23}&A_{24}&A_{25} \\
A_{31}&A_{32}&A_{33}&A_{34}&A_{35} \\
A_{41}&A_{42}&A_{43}&A_{44}&A_{45} \\
0     &A_{52}&0     &0     &0      \\
\end{array} \right)
\eqno{(A.2)}
$$
$$
A_{11}=v_xhn(1-c^2_s)+\frac{hv_x}{{\Gamma}^2}
$$
$$
A_{12}=\frac{1}{\Gamma}[n+v^2_x-nc^2_s(v^2_y+v^2_z)]
$$
$$
A_{13}=\frac{1}{\Gamma}v_xv_y(1+nc^2_s)
$$
$$
A_{14}=\frac{1}{\Gamma}v_xv_z(1+nc^2_s)
$$
$$
A_{15}=-\frac{1}{\Gamma}v_x(1+n)
$$
$$
A_{21}=-\frac{1}{\Gamma}(1-v^2_x)h^2(1-nc^2_s)
$$
$$
A_{22}=v_xh[2n(1-c^2_sv^2)-(1-v^2_x)(1+nc^2_s)]
$$
$$
A_{23}=-v_yh(1-v^2_x)(1+nc^2_s)
$$
$$
A_{24}=-v_zh(1-v^2_x)(1+nc^2_s)
$$
$$
A_{25}=-v^2_xh(1+n)+h(1+nc^2_sv^2)
$$
$$
A_{31}=\frac{1}{\Gamma}v_xv_yh^2(1-nc^2_s)
$$
$$
A_{32}=v_yh[n(1-c^2_sv^2)+v^2_x(1+nc^2_s)]
$$
$$
A_{33}=v_xh[n(1-c^2_sv^2)+v^2_y(1+nc^2_s)]
$$
$$
A_{34}=v_xv_yv_zh(1+nc^2_s)
$$
$$
A_{35}=-v_xv_yh(1+n)
$$
$$
A_{41}=\frac{1}{\Gamma}v_xv_zh^2(1-nc^2_s)
$$
$$
A_{42}=v_zh[n(1-c^2_sv^2)+v^2_x(1+nc^2_s)]
$$
$$
A_{43}=v_xv_yv_zh(1+nc^2_s)
$$
$$
A_{44}=v_xh[n(1-c^2_sv^2)+v^2_z(1+nc^2_s)]
$$
$$
A_{45}=-v_xv_zh(1+n)
$$
$$
A_{52}=hn(1-c^2_sv^2)
$$
$$
N=hn(1-c^2_sv^2)
$$

\clearpage

\begin{figure}
\vspace{-3cm}\hspace{0cm}\epsfxsize=16cm\epsfbox{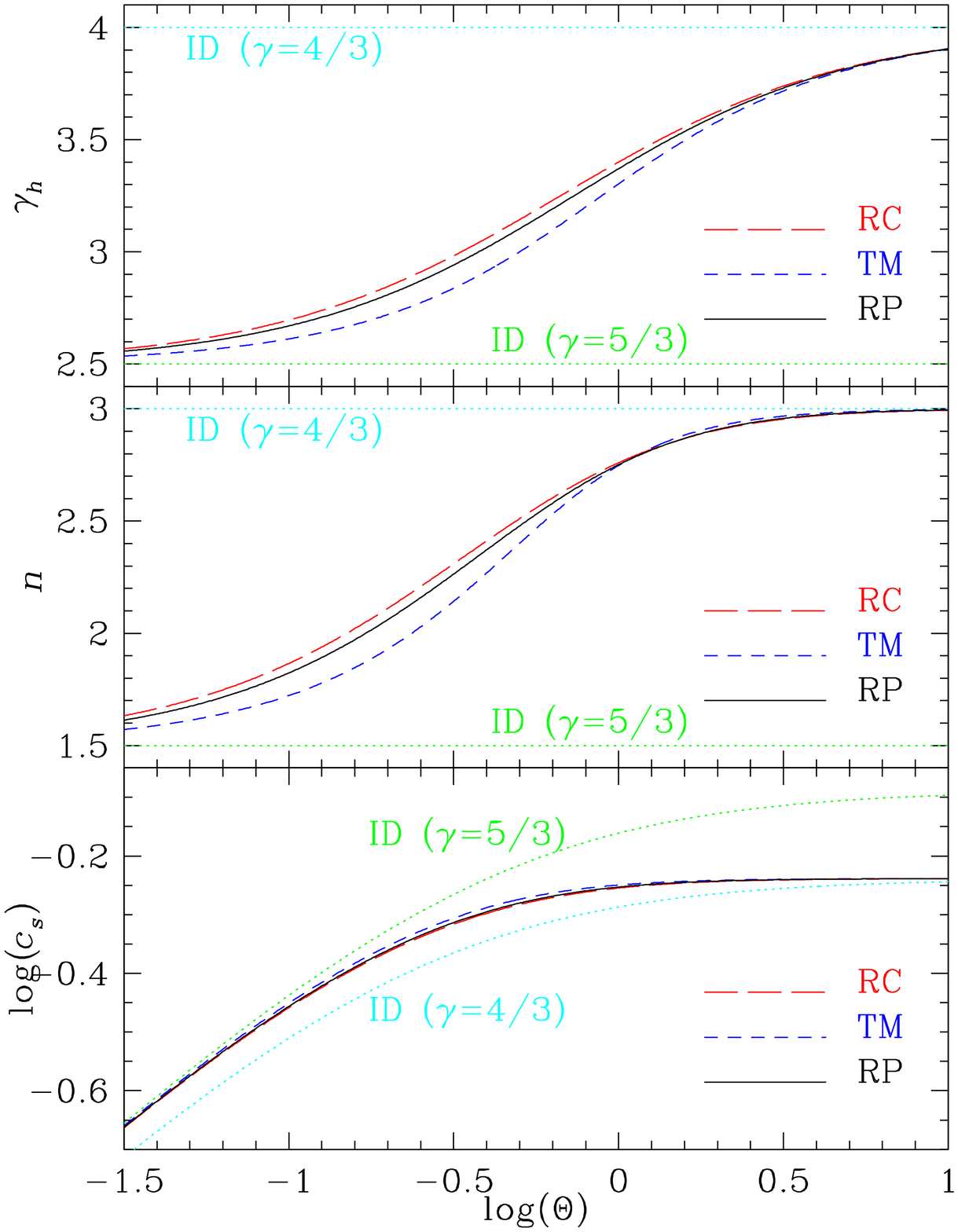}\vspace{-0.2cm}
\figcaption
{Comparison between different EoS's.
${\Gamma}_h$, $n$, and $c_s$, vs $\Theta$ for RC (red-long dashed),
TM (blue-short dashed), ID (green and cyan-dotted), and RP (black-solid).}
\end{figure}

\begin{figure}
\vspace{-4cm}\hspace{0cm}\epsfxsize=16cm\epsfbox{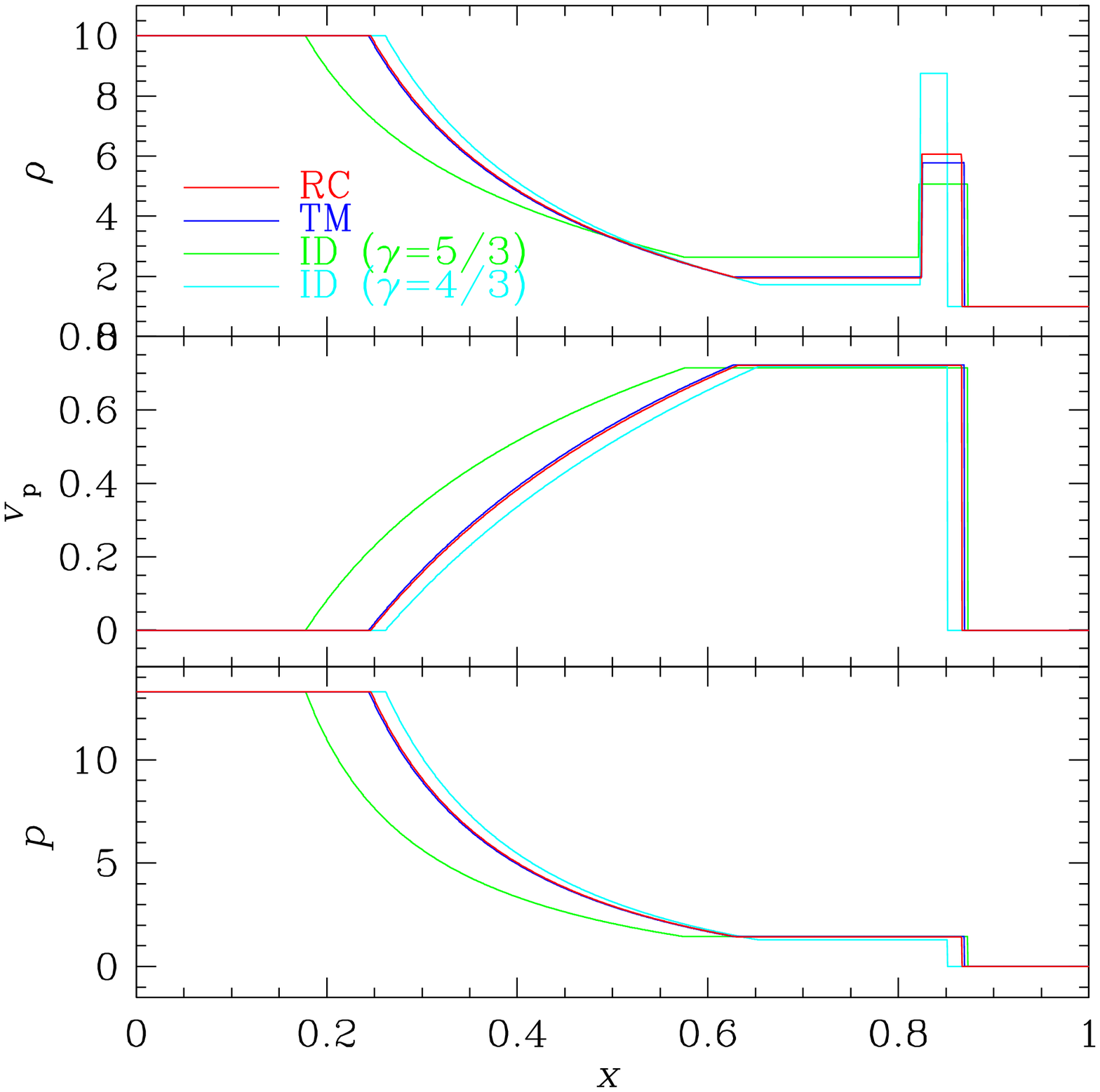}\vspace{-2.5cm}
\figcaption
{Relativistic shock tube with parallel component of velocity only
(P1) with RC (red), TM (blue), and ID (green and cyan).}
\end{figure}

\begin{figure}
\vspace{-4cm}\hspace{0cm}\epsfxsize=16cm\epsfbox{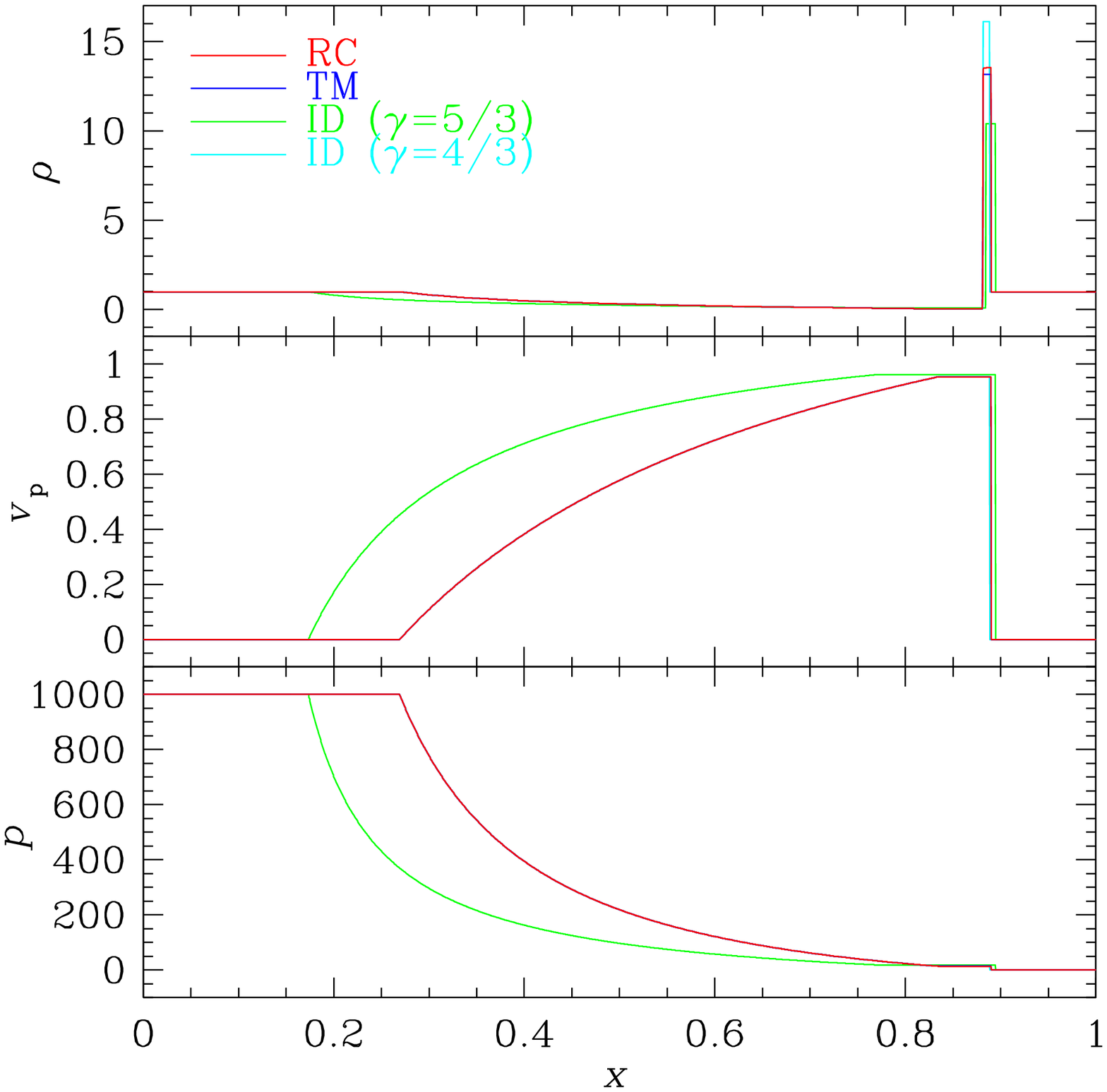}\vspace{-2.5cm}
\figcaption
{Relativistic shock tube with parallel component of velocity only
(P2) with RC (red), TM (blue), and ID (green and cyan).}
\end{figure}

\begin{figure}
\vspace{-3cm}\hspace{0cm}\epsfxsize=16cm\epsfbox{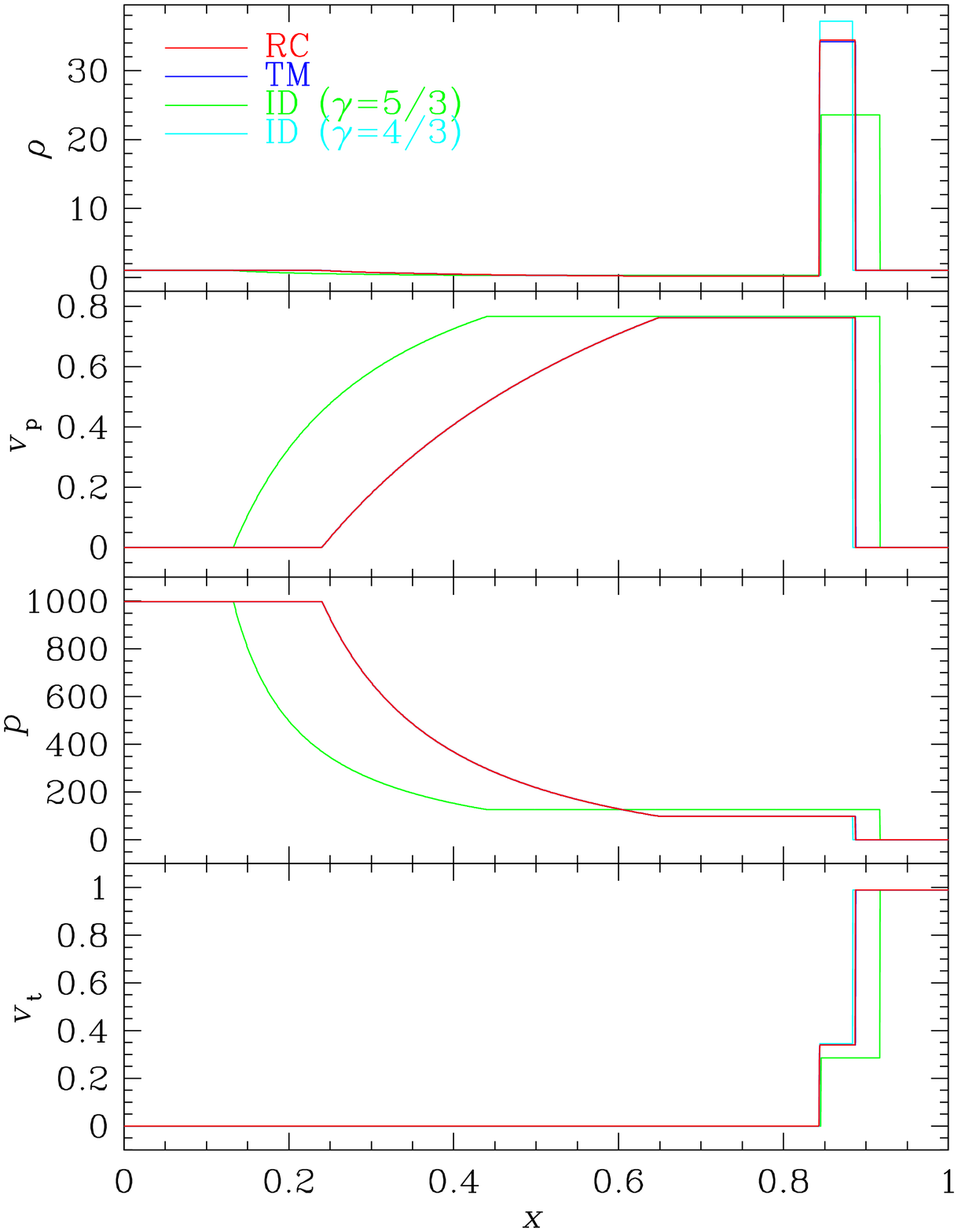}\vspace{-0.2cm}
\figcaption
{Relativistic shock tube with transverse component of velocity (T1)
with RC (red), TM (blue), and ID (green and cyan).}
\end{figure}

\begin{figure}
\vspace{-3cm}\hspace{0cm}\epsfxsize=16cm\epsfbox{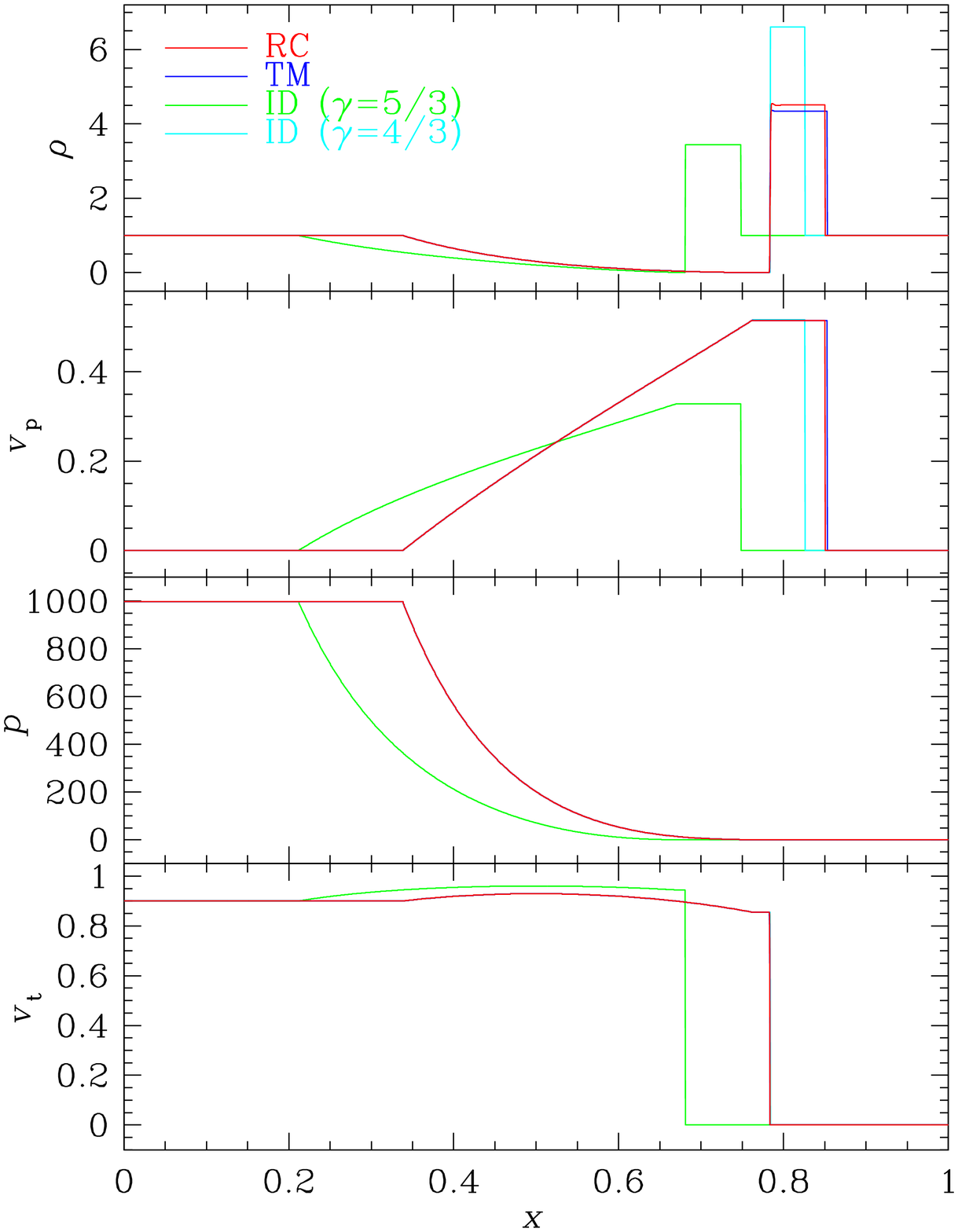}\vspace{-0.2cm}
\figcaption
{Relativistic shock tube with transverse component of velocity (T2)
with RC (red), TM (blue), and ID (green and cyan).}
\end{figure}

\begin{figure}
\vspace{-3cm}\hspace{0cm}\epsfxsize=16cm\epsfbox{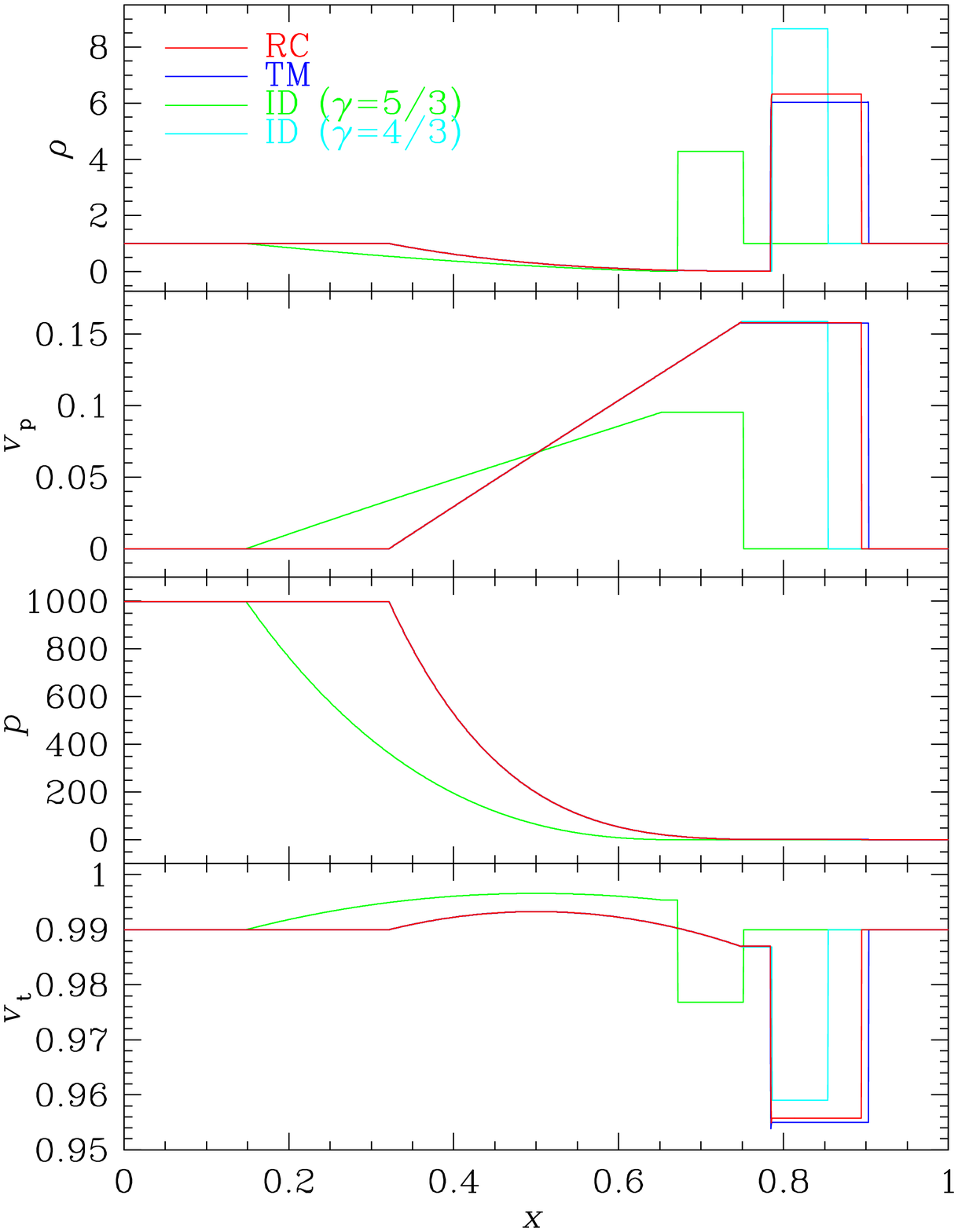}\vspace{-0.2cm}
\figcaption
{Relativistic shock tube with transverse component of velocity (T3)
with RC (red), TM (blue), and ID (green and cyan).}
\end{figure}

\begin{figure}
\vspace{-3cm}\hspace{0cm}\epsfxsize=16cm\epsfbox{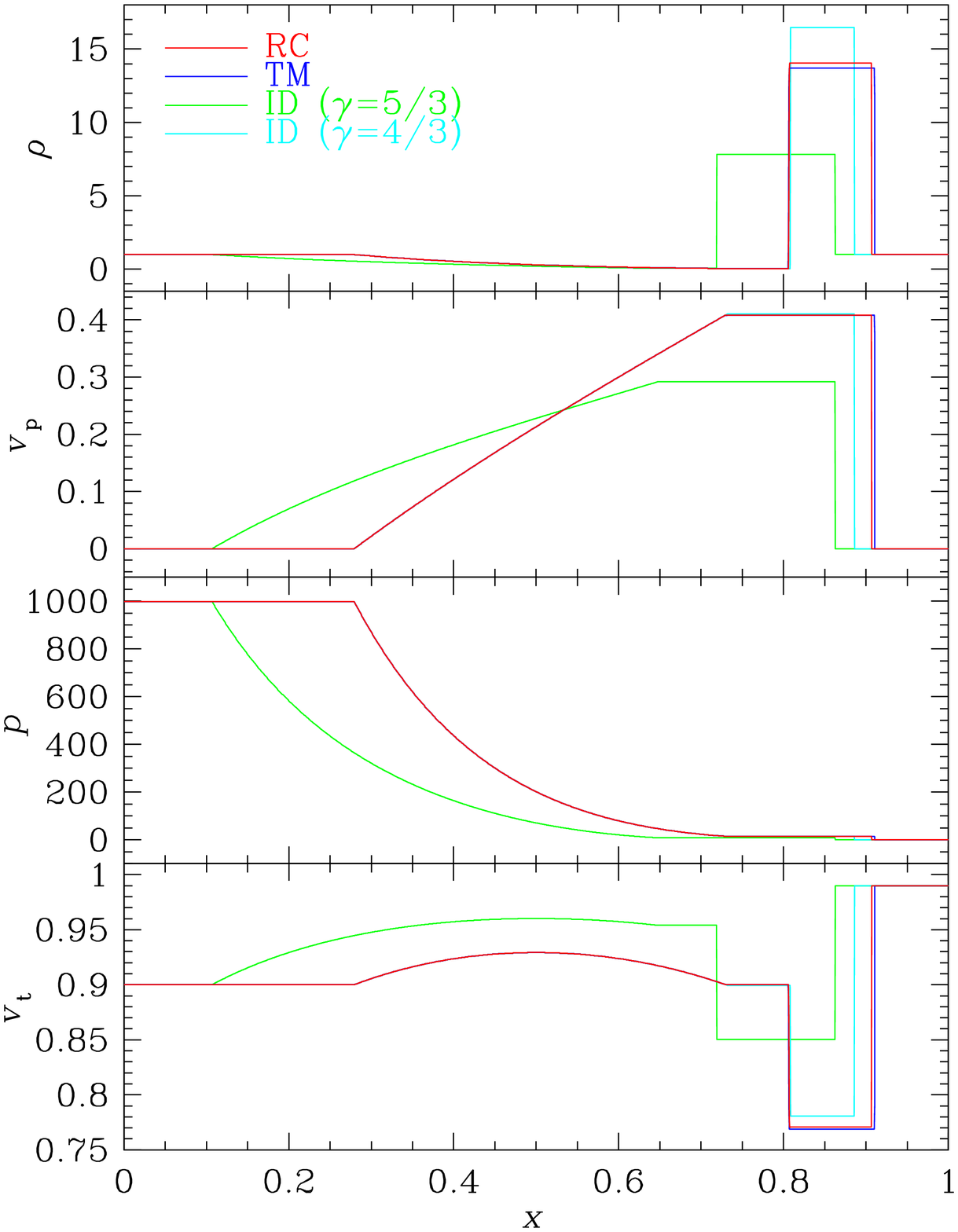}\vspace{-0.2cm}
\figcaption
{Relativistic shock tube with transverse component of velocity (T4)
with RC (red), TM (blue), and ID (green and cyan).}
\end{figure}

\end{document}